# Associative learning in biochemical networks

Nikhil Ghandi
*College of Computing, Georgia Institute of Technology, Atlanta, GA 30332*

Gonen Ashkenasy[1]
*Department of Chemistry, Ben-Gurion University of the Negev, Be'er-Sheva 84105, Israel*

Emmanuel Tannenbaum[2]
*School of Biology, Georgia Institute of Technology, Atlanta, GA 30332*

**Abstract**

It has been recently suggested that there are likely generic features characterizing the emergence of systems constructed from the self-organization of self-replicating agents acting under one or more selection pressures. Therefore, structures and behaviors at one length scale may be used to infer analogous structures and behaviors at other length scales. Motivated by this suggestion, we seek to characterize various "animate" behaviors in biochemical networks. Specifically, in this paper, we develop a simple, chemostat-based model illustrating how a process analogous to associative learning can occur in a biochemical network. Associative learning is a form of learning whereby a system "learns" to associate two stimuli with one another. Associative learning, also known as conditioning, is believed to be a powerful learning process at work in the brain (associative learning is essentially "learning by analogy"). In our model, two types of replicating molecules, denoted A and B, are present in some initial concentration in the chemostat. Molecules A and B are stimulated to replicate by some growth factors, denoted $G_A$ and $G_B$, respectively. It is also assumed that A and B can covalently link, and that the conjugated molecule can be stimulated by either the $G_A$ or $G_B$ growth factors (and can be degraded). We show that, if the chemostat is stimulated by both growth factors for a certain time, followed by a time gap during which the chemostat is not stimulated at all, and if the chemostat is then stimulated again by only one of the growth factors, then there will be a transient increase in the number of molecules activated by the other growth factor. Therefore, the chemostat bears the imprint of earlier, simultaneous stimulation with both growth factors, which is indicative of associative learning. It is interesting to note that the dynamics of our model is consistent with various aspects of Pavlov's original series of associative learning experiments in dogs. We discuss how associative learning can potentially be performed *in vitro* within RNA, DNA, or peptide networks. We also highlight how such a mechanism could potentially be involved in genomic evolution, and suggest bioinformatics studies that could be used to find evidence for associative learning processes at work inside living cells.

# I. Introduction

Emerging evidence suggests that much of the so-called "junk" DNA in complex, multi-cellular eukaryotic organisms in fact codes for a vast, RNA-based, genetic regulatory network (Claverie, 2005; Costa, 2005; Dennis and Omer, 2005; Green and Doudna, 2006; Herbert, 2004; Herbert and Rich, 1999; Laaberki and Repoila, 2003; Mattick, 2005; Mattick and Makunin, 2006; Moulton, 2005; Plasterk, 2006; Wassarman, 2004). In addition to the regulatory roles in encoding protein structure, RNA is involved in other processes such as gene silencing, catalysis of chemical reactions, and sensing self and non-self analytes. It is believed that this RNA biochemistry is responsible for the variety and complexity of terrestrial life. Since the RNA biochemistry emerged through a long process of replicative selection, it is likely that there are large subnetworks of RNA interactions that are essentially biochemical implementations of some fairly sophisticated computational algorithms associated with proper cell function. Therefore, a major challenge for systems and evolutionary biologists will be to uncover the structure of these biochemical networks, and understand their evolution and the role they play in the emergence of complex terrestrial life.

In a recent paper (Tannenbaum, 2006), Tannenbaum argued that the hypothesized RNA networks in complex eukaryotes exhibit structures and behaviors that are analogous to structures and behaviors that emerge in agent-built systems such as the brain (it is believed that pathway selection in the brain is driven by a chemically-based reward-punishment system). This speculation was driven by the hypothesis that there are likely generic features in systems that are constructed by agents acting under one or more selection pressures (replicative selection, reward-punishment chemicals in the brain, free-market competition, etc.). Therefore, by studying structures and behaviors at one length scale, it may be possible to infer the existence of analogous structures and behaviors at another scale.

Because of the sheer magnitude and complexity of the hypothesized RNA biochemistry in eukaryotic cells, determining the various RNA pathways will be extremely difficult without *a priori* guesses as to what kinds of structures to look for. Furthermore, even if one constructs a detailed map of the various RNA networks, such a map will provide almost no insight into the logic of the network. That is, a detailed representation of the RNA biochemistry inside the cell will simply appear to be a massively complex, but largely random, web of RNA interactions.

However, if there are indeed generic features associated with agent-built systems acting under various selection pressures, then by observing structures such as the brain, it may be possible to infer the existence of analogous structures inside RNA biochemical networks.

Based on the reasoning presented here, we seek to elucidate the various "animate" behaviors (or computational motifs) implemented by the RNA biochemistry inside living cells. In this paper specifically, we propose the existence of an RNA-based implementation of a computational scheme known as *associative learning* (Kohonen, 1989; Mackintosh, 1983). Associative learning is believed to be a key aspect of the thought processes at work in the brain (Kohonen, 1989; Mackintosh, 1983). While these concepts will be discussed in more detail in the paper, associative learning may be briefly described as a form of learning whereby various stimuli become associated with one another.

Based on what is currently known about the RNA biochemistry at work inside eukaryotic cells, we argue that the proposed RNA-based associative learning scheme occurs in eukaryotes, and may even lead to genomic evolution. As support for this argument, we develop in this paper a "toy" model that illustrates how associative learning could potentially occur in a biochemical network. We also discuss future *in vitro* experiments that could be used to establish the realisability of this computational motif. Finally, we also discuss some possible bioinformatics studies that could be used to search for evidence that associative learning processes have played roles in genomic evolution.

## II. A Minimalist Model for Associative Memory

## A. Definitions

We begin by defining the various concepts used in this paper, namely, memory, learning, associative memory, and associative learning.

*Memory* in a physical system refers to the ability of the system to preserve information about the state of the universe (including the system itself) at some previous time. Simple examples include childhood recollections (first day of school, first bike, etc.) (Kohonen, 1989). An important biological example is the immune system: After the immune system fights off an infection, a fraction of the antibody-producing B-cells turn into memory cells, that allow the immune system to rapidly respond to future invasions by a given antigen.

*Learning* refers to the ability of a system to acquire new functions in response to some external input. Simple examples include learning how to read, cook, or play a new sport (Vapnik, 1998). Learning is similar to the concept of adaptation. Strictly speaking, learning refers to the acquisition of new functions, while adaptation refers to changes in a system that allow it to function in a new environment. However, because adaptation can occur via the acquisition of new functions, then, with an appropriate definition of the system, the concepts of learning and adaptation can be shown to be formally equivalent.

The immune response to a new infectious agent is an important example of learning exhibited at the cellular level. The population of antibody-producing B-cells evolves through a process of clonal selection and somatic hypermutation, until it becomes optimally tailored to fight the infectious agent. Clonal selection and somatic hypermutation are analogous to learning by trial-and-error, whereby the immune system tests various antibodies against a given antigen. Those antibody designs that are most effective within the given antibody population are used as templates for further refinements, so that, after several iterations, the optimal antibody design is found.

It is believed that pathway selection in the brain occurs via processes that are analogous in many respects to the learning processes associated with the immune response (Ashton et al., 2002).

*Associative memory* is a form of memory whereby the stimulation of one memory triggers the stimulation of another. The two separate memories are essentially components of a larger, compound memory, and so stimulating one component of the compound memory stimulates the whole memory.

*Associative learning* is a learning process where two or more distinct stimuli become associated with one another. In a sense, then, associative learning refers to the process by which an associative memory is created.

A famous 19th century example of associative learning at work in the brain is the series of experiments by Ivan Pavlov, in which a dog was simultaneously stimulated with the sound of a ringing bell and the sight of food. The sight of food caused the dog to salivate. Eventually, the dog would salivate from the sound of the ringing bell alone.

In a biochemical network, a signature of memory may be the production of a certain compound as a result of a certain input into a system. A signature of associative memory and/or learning is then the production of distinct compounds, each having a separate external stimulus, as a result of input of the stimulus for only one of the compounds. In such a case, the system behaves at if it was stimulated with several inputs, which are therefore effectively associated with one another.

**B. Kinetic Model**

To develop a simple model that can exhibit associative learning, we consider a chemostat of volume V, containing two polynucleotide species: (1) Species A, characterized by some base sequence $\sigma_A$, is stimulated to replicate via some growth factor $G_A$. (2) Species B, characterized by some base sequence $\sigma_B$, is stimulated to replicate via some growth factor $G_B$. Pictorially, these two reactions may be represented via,

$$A + G_A \rightarrow A + A$$
$$B + G_B \rightarrow B + B \tag{1}$$

where we assume that the reaction kinetics are second-order, with a species-independent rate constant $k_R$ (in the context of a genome, the growth factors $G_A$ and $G_B$ correspond to transcription factors that trigger RNA production from a given DNA polynucleotide sequence).

We assume that the species A and B can chemically react, to form either the chain $\sigma_A$-$\sigma_B$ or $\sigma_B$-$\sigma_A$, both of which may be termed species A – B. We also assume that A - B can dissociate, so that the forward and back reactions are given by,

$$A + B \leftrightarrow A - B \tag{2}$$

where the forward reaction has a second-order rate constant $k_f$, and the back reaction has a first-order rate constant $k_b$.

Finally, we assume that the growth factors $G_A$ and $G_B$ can both stimulate replication of A-B, via the reaction,

$$A - B + G_{A/B} \rightarrow A - B + A - B \tag{3}$$

which also is assumed to proceed with a second-order rate constant $k_R$. As will be seen, the assumption that the replication of A-B can be stimulated by either growth factor is the key assumption in our model, and may be understood as follows: As mentioned previously, $G_A$ and $G_B$ may both be thought of as analogous to transcription factors, that essentially "unlock" the promoter region of a polynucleotide sequence and allow the replicase enzyme to bind and catalyze replication. The molecule A-B has two promoter regions, each of which can be "unlocked" by their respective transcription factors. We are assuming that unlocking the entire molecule (for access by the replicase) is achievable by unlocking the molecule at one site (as a useful analogy, imagine a cylinder that is capped at both ends. Uncapping one of the ends unlocks the whole cylinder).

We assume a volumetric flow rate F through the chemostat, and that the input concentration of $G_A$ and $G_B$ is given by $c_{A,0}$ and $c_{B,0}$, respectively. If $n_{G_A}, n_{G_B}, n_A, n_B, n_{AB}$ denote the population numbers of $G_A$, $G_B$, A, B, and A - B, respectively, then we have the following system of differential equations governing the chemical reaction kinetics inside the chemostat:

$$\frac{dn_{G_A}}{dt} = Fc_{A,0} - \frac{k_R}{V}(n_A + n_{AB})n_{G_A} - \frac{F}{V}n_{G_A}$$

$$\frac{dn_{G_B}}{dt} = Fc_{B,0} - \frac{k_R}{V}(n_B + n_{AB})n_{G_B} - \frac{F}{V}n_{G_B}$$

$$\frac{dn_A}{dt} = \frac{k_R}{V}n_A n_{G_A} - \frac{F}{V}n_A - \frac{k_f}{V}n_A n_B + k_b n_{AB}$$

$$\frac{dn_B}{dt} = \frac{k_R}{V}n_B n_{G_B} - \frac{F}{V}n_B - \frac{k_f}{V}n_A n_B + k_b n_{AB}$$

$$\frac{dn_{AB}}{dt} = \frac{k_R}{V}n_{AB}(n_{G_A} + n_{G_B}) - \frac{F}{V}n_{AB}$$

$$+ \frac{k_f}{V}n_A n_B - k_b n_{AB}$$

(4)

To simplify these equations, we assume that $k_f$ and $k_b$ are sufficiently large that the reaction $A + B \leftrightarrow A - B$ is always in equilibrium. We let $K = k_f/k_b$ denote the equilibrium constant. We also define $\tilde{n}_A = n_A + n_{AB}$, and $\tilde{n}_B = n_B + n_{AB}$. Finally, we define $f_A = Fc_{A,0}$, $f_B = Fc_{B,0}$, and $f = F/V$. Putting everything together, we obtain,

$$\frac{dn_{G_A}}{dt} = f_A - \frac{k_R}{V}\tilde{n}_A n_{G_A} - fn_{G_A}$$

$$\frac{dn_{G_B}}{dt} = f_B - \frac{k_R}{V}\tilde{n}_B n_{G_B} - fn_{G_B}$$

$$\frac{d\tilde{n}_A}{dt} = \frac{k_R}{V}\tilde{n}_A n_{G_A} + \frac{k_R}{V}n_{AB}(\tilde{n}_A, \tilde{n}_B)n_{G_B} - f\tilde{n}_A$$

$$\frac{d\tilde{n}_B}{dt} = \frac{k_R}{V}\tilde{n}_B n_{G_B} + \frac{k_R}{V}n_{AB}(\tilde{n}_A, \tilde{n}_B)n_{G_A} - f\tilde{n}_B$$

(5)

where,

$$n_{AB}(\tilde{n}_A, \tilde{n}_B) = \frac{1}{2}[\tilde{n}_A + \tilde{n}_B + \frac{1}{K} - \sqrt{(\tilde{n}_A + \tilde{n}_B + \frac{1}{K})^2 - 4\tilde{n}_A \tilde{n}_B}] \quad (6)$$

To see how associative learning emerges from the chemostat dynamics, we consider the following experiment: Starting with some initial seed population of A and B molecules inside the tank, the system is fed with growth factors $G_A$ and $G_B$ at rates $f_{A,1}$ and $f_{B,1}$. This proceeds for some time until the system has reached a steady-state. While maintaining a constant volumetric flow rate through the system (so that f remains unchanged), $f_A$ and $f_B$ are rapidly brought to 0, and are maintained at 0 for a certain period of time, denoted T.

After this time T, $f_A$ is raised to some value $f_{A,2}$, while $f_B$ is maintained at 0. In this second regime, it is possible to show that the steady-state value of $\tilde{n}_B$ is 0. To see why, note that the steady-state value of $n_{G_B}$, denoted $n_{G_B,ss}$, is 0. Therefore, the differential equation for $n_A$ gives, at steady-state,

$$\frac{k_R}{V} n_{G_A,ss} = f \quad (7)$$

so that the steady-state equation for $\tilde{n}_B$ becomes,

$$0 = -f(\tilde{n}_{B,ss} - n_{AB,ss}) = -fn_{B,ss} \quad (8)$$

which gives $n_{B,ss} = 0$. Coupled to the equilibrium criterion, $n_{AB} = K n_A n_B$, we obtain $n_{AB,ss} = 0$, and hence $\tilde{n}_{B,ss} = 0$.

Therefore, after a sufficiently long amount of time, stimulation of the chemostat by $G_A$ alone will eventually lead to the disappearance of all B molecules from the chemostat.

However, the transient behavior of the dynamics is qualitatively different for the $K = 0$ and $K > 0$ cases. When $K = 0$, then $n_{AB} = 0$, and so the dynamical equation for $\tilde{n}_B$ is,

$$\frac{d\tilde{n}_B}{dt} = \tilde{n}_B(\frac{k_R}{V} n_{G_B} - f) \quad (9)$$

If the time gap T is sufficiently large, then $n_{G_B}$ will be sufficiently small so that $\frac{k_R}{V} n_{G_B} - f$ and so $\tilde{n}_B$ will steadily decrease with time.

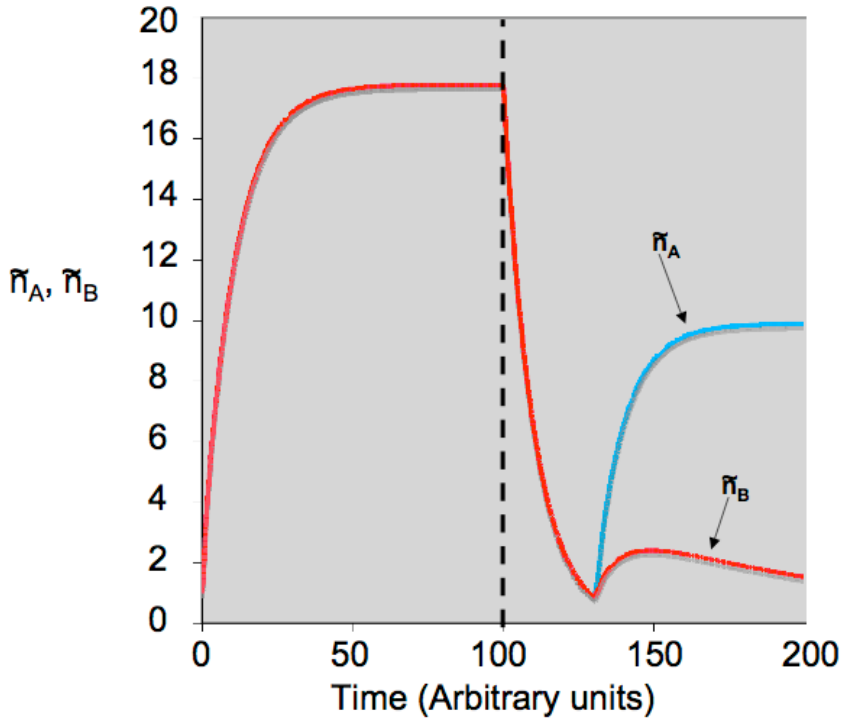

**Figure 1:** Illustration of associative learning in our toy model (Eq. (5)). The system is first stimulated with both $G_A$ and $G_B$ until some time $T_1 = 100$, at which point the system is at steady-state. The flow of $G_A$ and $G_B$ is then turned off from $T_1$ to $T_2 = 130$, so that the time gap $T = 30$. At $T_2 = 130$, the system is again stimulated, but only with $G_A$. The value of $\tilde{n}_A$ rises to a steady-state value, while the value of $\tilde{n}_B$ has a transient increase, but then begins to slowly decrease to 0. Parameter values are $f = 0.1$, $f_{A,1} = f_{A,2} = f_{B,1} = 1$, $k_R/V = 1$, and $K = 1$.

However, when $K > 0$, then the presence of $n_{AB}$ in the population can lead to an initial, transient increase in the value of $\tilde{n}_B$ after $G_A$ is again allowed to flow into the system (the larger the value of K, the larger and more persistent the increase). This transient increase in the number of B molecules in the solution is exactly indicative of associative learning, because it results from the covalent linkage of A and B molecules previously generated by simultaneously feeding the system both $G_A$ and $G_B$. The population in the chemostat has essentially "learned" to associate $G_A$ and $G_B$, in the sense that stimulation with only $G_A$ at some future time leads to a transient signature of stimulation with $G_B$ as well. Figure 2 provides an example clearly illustrating this phenomenon.

It should be noted that the transient nature of the association means that, if the system is stimulated by $G_A$ alone, then, the system will eventually "forget" the earlier association between $G_A$ and $G_B$. This is consistent with the original experiments by Pavlov: If the bell was rung a sufficient number of times without subsequently presenting the dog with food, the association between the ringing bell and the food was lost, and the dog would stop salivating at the sound of the ringing bell alone.

## C. Possible *in vitro* Experimental Tests

The simplest biochemical networks can be constructed from RNA (Kim and Joyce, 2004), DNA (Sievers and von Kiedrowski, 1994; Sievers and von Kiedrowski, 1998) or peptide (Ashkenasy et al., 2004; Lee et al., 1997b; Yao et al., 1998) molecules that exploit non-enzymatic replication as recurring elements (Lee et al., 1997a; Paul and Joyce, 2004). It was shown recently that modules of such networks perform (simple) computational algorithms such as reciprocal replication (Ashkenasy et al., 2004; Kim and Joyce, 2004; Sievers and von Kiedrowski, 1998) and Boolean logic functions (Ashkenasy et al., 2004). The molecular setup described below is customized to include the different steps of the associative memory algorithm. The possibility of the suggested system to actually perform the associative learning is high, since all the aspects of this process have been demonstrated with similar chemical entities (see references below). In the near future we intend to prepare the building blocks and characterize such systems experimentally.

    In our experimental setup, the self-replication process of A and B molecules requires two steps (a and b in Figure 2; shown only for A molecules). The first is a reversible initiation step by which the replicating molecule is activated by an external trigger. It is shown schematically as a cleavage process, equivalent to enzyme activation through dephosphorylation or photocleavage, but can represent in a more general sense any molecular-activating conformational changes. In the second step, the activated molecules A* and B* serve as templates for association and ligation of their own fragments (e.g. $A_1$ and $A_2$ that produce A). Molecules A and B can then bind to each other to form the long polymer (step c). To account for reversible binding between A and B we consider backbone modification - usually manifested in nucleic acids or peptides through ester or thioester formation. The replication of the A-B polymer can be triggered through activating of molecule A alone if we consider a scenario by which the template-product complex (A*-A) can serve as template for the formation of B molecules, a process that will lead to high local concentration of A and B and as a result to the formation of more A-B molecules. Molecular replication processes that rely on high oligomerization states (i.e. trimers or tetramers) have been observed both in the DNA and peptide based systems, suggesting that the postulated step is plausible.

    The kinetic experiment will be performed by placing together molecules A and B and their precursor $A_1$, $A_2$ and $B_1$, $B_2$ respectively, and initiating replication through introduction of external triggers (e.g. shining light). After a time that corresponds to significant formation of additional A and B molecules through self-replication, and probably also through reciprocal replication (step d of Figure 2), the external triggering will be turned off and the system will be allowed to equilibrate. Since all the processes are reversible, it is expected that after some time the system will reach equilibrium with a smaller, but nonzero, amount of the full length molecules A, B and A-B present. Applying the trigger that activates only the A molecules will start a cascade reaction that will increase the amount of A itself, but more importantly that of the A-B and thus also of the B molecules.

    As explained above, the requirements for replication of B in the absence of its own trigger, what we called associative memory or learning, is the existence of some

(even small) amounts of B molecules in the mixture, and not of less importance the ability of A to activate B replication as a consequence of the formation of A-B molecules.

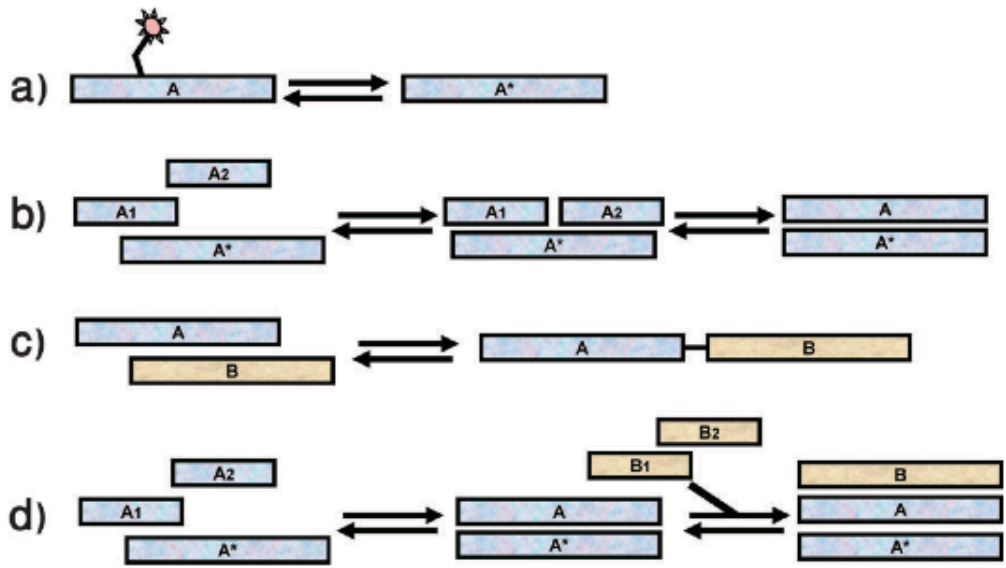

**Figure 2:** Associative learning process within a simple biochemical network. Molecules A and B represent either DNA, RNA, or polypeptides. The reversible chemical reactions in a through d show the kinetic steps of the process under study as explained in the text.

### III. Discussion

## A. How Associative Memory and Learning Speed Up the Rate of Adaptation

Associative memory and learning can drastically speed up rates of adaptation by triggering a system to use the solution of one problem to solve a new, but similar problem. To understand this, consider first the example of how someone who knows how to surf might find it easier to learn how to snowboard than someone starting for the first time.

The person who knows how to surf needed to learn how to surf at one time. The process of surfing has several aspects, or properties, each of which may be regarded as separate inputs into the brain: (1) The shape of the board. (2) The body position when standing on the board. (3) The body positions one must assume to maintain stability while surfing a wave. (4) The speed associated with surfing.

These separate but simultaneous inputs become associated with each other in the brain, so that stimulation of one of the memories generated by these inputs may stimulate the others. So, it is likely that a surfer who starts learning how to snowboard will have the memory of the surfboard triggered by the sight of the snowboard. This in turn will trigger the memories of the various body positions and movements associated with surfing, which may then form the basis for learning the body positions and movements for snowboarding.

## B. Biochemical Implementations of Associative Learning

Associative learning in a eukaryotic cell could occur via a slightly different mechanism than the simple chemostat model described in this paper. In a eukaryotic cell, two distinct transcription factors could lead to transcription of two distinct genes, resulting in a steady production of two distinct sets of RNAs (these RNAs could be mRNAs, or, in the case of the non-coding regions, these could be various RNAs that are never translated into protein, such as siRNA). If some of these RNAs were to covalently bind, and if these longer RNAs were then reverse transcribed into the DNA genome, the result would be a new DNA sequence that corresponds to the two original genes in succession. If either of the original transcription factors could then activate this new gene set, then the end result would exactly be a DNA-based implementation of associative learning. This scheme is illustrated in Figure 3.

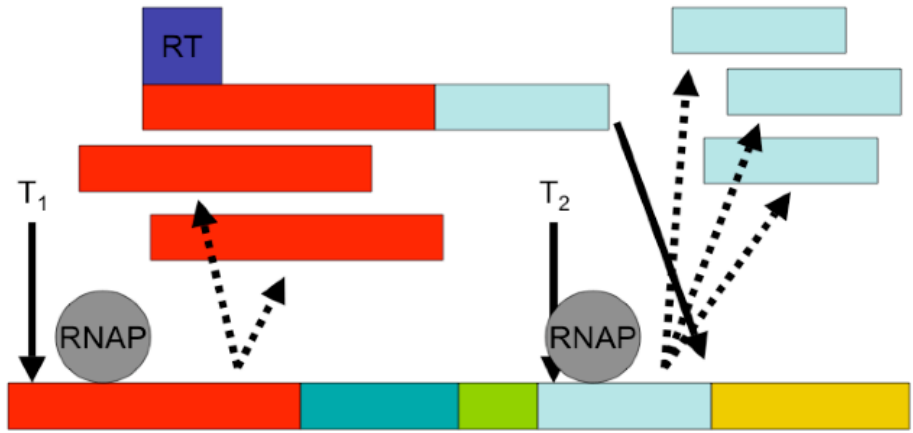

**Figure 3:** Implementation of associative learning in a eukaryotic cell. Two RNA transcripts from two distinct genes become covalently linked. Reverse transcription then results in two distinct genes becoming linked in the genome.

The argument presented here of course assumes that reverse transcriptase is active in eukaryotic cells. In support of this claim, it is believed that gene duplication in eukaryotes occurs primarily through a mechanism known as retrotransposition, whereby the gene first goes through an RNA intermediate that is then reverse transcribed into the DNA genome. Recent work and speculation has also suggested that much genomic change in eukaryotes first occurs in the RNA population of the cell, and only then is later reverse transcribed into the genome (Herbert and Rich, 1999). The key role that the RNA population inside eukaryotic cells is hypothesized to play in genomic evolution has led to the term *ribotype*, as a way of characterizing the distribution of RNA sequence types (Herbert and Rich, 1999).

## C. Finding Evidence of Associative Learning in Actual Genomes

The strongest evidence for associative learning in biochemical systems would be its actual discovery in real organisms. Therefore, an essential line of research would be to look for certain genes or genome regions in eukaryotic cells that appear to be linkages of other, smaller genes or genome regions.

One possible example of associative learning processes in genomes is the existence of polycistronic RNA in prokaryotic organisms. Because proteins generally function as part of interconnected biochemical networks, when one protein needs to be produced by a cell, generally several other proteins need to be produced as well.

Polycistronic RNA in bacteria is an mRNA transcript that encodes for several proteins. Basically, when one gene encoding for part of a biochemical network is transcribed, the other genes coding for the remainder of the network are transcribed as well.

Although polycistronic RNA occurs in prokaryotes, and not eukaryotes, which are the focus of this paper, it is believed that prokaryotes and eukaryotes evolved simultaneously from the archaebacteria, whose genome organization is more similar to that of the eukaryotes than that of the prokaryotes. Therefore, it is possible that polycistronic RNA evolved from the linking and then reverse transcription of various mRNA transcripts in archaebacterial cells. These mRNA transcripts were present at the same time because they all encoded for essential pieces of some biochemical network, and so their production was triggered by the presence of several transcription factors. If these mRNA transcripts then became covalently linked, and were then reverse transcribed back into the archaebacterial genome, the result would be a sequence of genes encoding for a polycistronic mRNA transcript.

## IV. Concluding Remarks

This paper presented a simple, "toy" model illustrating how associative learning could occur in a biochemical network. We discussed how this process could be implemented in eukaryotic cells, and how it could lead to genomic evolution.

The model presented in this paper only shows that associative learning is in principle possible inside living cells. Experiments and bioinformatics studies will be needed to confirm (or disprove) the existence of associative processes at work inside living cells.

We should point out that the concept of associative processes in polypeptides or polynucleotides has been considered before. In 1995, Eric B. Baum developed a scheme for using double-stranded DNA to construct a large associative memory (Baum, 1995). However, the associative memory considered by Baum was for the purpose of retrieving information in a database encoded in the DNA chain. The idea is that incomplete information, in the form of short strands of DNA, could attach to corresponding subsequences along the DNA, and thereby retrieve all possibly relevant database items (Baum, 1995).

This paper, by contrast, considers an associative process generated by stimulating a system with two simultaneous inputs. The association of the two inputs is not "hard-wired" as in the model considered by Baum, but rather is learned (a DNA-based memory scheme closer to the model considered in this paper may be found in (Chen et al., 2005)). Both types of associative processes may be at work inside eukaryotic cells. Nevertheless, we conjecture that, if associative processes are indeed a major mechanism for macro-evolutionary change, then it is likely via a mechanism closer to the one outlined here. This speculation is derived from the observation that species evolve in response to external selection pressures. Therefore, if associative processes are indeed relevant to macroevolution, then it appears more likely to be a form of associative learning, whereby external inputs drive the associations produced.

If associative processes analogous to the one presented here are found to occur in living cells, then our hypothesis that there are likely generic features in agent-built systems may indeed be correct. Certainly, the existence of associative processes inside living cells would suggest the need for further studies to elucidate other computational structures at work inside living cells (Landweber et al., 2000), and the possible role these computational structures play in genomic evolution.

## Acknowledgments

The authors thank Mark Borodovsky (Georgia Tech) for helpful conversations regarding this work. GA thanks the Human Frontier Science Program for Career Development Award.